\documentstyle[seceq,supplement,epsf,float,mbf]{ptptex}

\newcommand{\bzero}{{\bf 0}}
\newcommand{\bq}{{\mbf q}}
\newcommand{\bk}{{\mbf k}}
\newcommand{\bv}{{\mbf v}}

\newcommand{\Imag}{{\rm Im}}

\title{Soft Modes at the Critical End Point \\ in the Chiral Effective Models}

\author{H.~{\sc Fujii}$^a$ and M.~{\sc Ohtani}$^b$}

\inst{
$^a$Institute of Physics, University of Tokyo, Tokyo 153-8902\\
$^b$Radiation Laboratory, RIKEN,  Saitama 351-0198}

\recdate{
\today
}

\abst{
At the critical end point in QCD phase diagram,
the scalar, vector
and entropy susceptibilities are known to diverge.
The dynamic origin of this divergence is identified
within the chiral effective models as softening
of a hydrodynamic mode of the particle-hole--type
motion, which is a  consequence of the conservation
law of the baryon number and the energy.
}

\begin{document}

\maketitle

\section{Introduction and Summary}

Theoretical studies of QCD 
thermodynamics have been predicting  rich phase structure
in the plane of the temperature $T$ and the chemical potential
$3\mu$
for baryon number, as discussed in the workshop,
and there are heavy--ion programs now ongoing and planned
in order to explore these new states of matter experimentally.
Based on the lattice QCD results\cite{FK02}
as well as the effective model calculations%
\cite{AY89,GGP94,BR99,HJSSV98}
it is strongly argued that
as one of the prominent features of the real diagram
 there exists a critical end point (CEP)%
\footnote{In condensed matter physics
the term {\em critical end point} is used to indicate a different
point where a critical line is truncated by meeting a first--order
phase boundary.}
 of
the first--order transition line, which is reviewed
by M.A.~Stephanov\cite{MS03}. Since it is a truly singular point,
we expect clearer information of it from experiments.
The basic property of CEP 
is the divergence of the susceptibilities
of the baryon number and entropy densities as well as
of the scalar density. 
Thus it is suggested that 
the fluctuations in the low--momentum
pions and nucleons show anomalous behavior as functions
of experimental parameters 
near CEP\cite{RS98,RS99}.
Theoretical estimates
indicate the possibility 
that the CEP gives still significant effects 
on these observables 
although the finiteness of the experimental geometry in space and time
will round off the critical 
growth of these 
fluctuations\cite{RS99,BR00}.

Near the critical point there will be at least one mode
which becomes soft representing the criticality\cite{HH77}.
In this report
we address the issue of the soft mode associated with CEP,
which causes these critical divergences,
within the Nambu--Jona-Lasinio model (NJL)\cite{HF03,FO03} and
the time--dependent Ginzburg--Landau (TDGL) approach\cite{FO03}.
The correct 
identification of the soft mode
associated with CEP is crucial 
to determine 
the time evolution of the critical fluctuations
and to study the experimental signals.

So far the static properties of CEP have been investigated,
where the underlying (approximate)
chiral symmetry is usually emphasized.
The effective potential 
with the scalar density $\sigma$ as the order parameter,
becomes flat
around a nonzero equilibrium value of $\sigma$ at CEP,
reflecting the divergence of the scalar susceptibility.
This seemingly suggests the appearance of the gapless
sigma mode there.
We note here, however,
that the baryon number and entropy susceptibilities
diverge at CEP simultaneously.
The scalar fluctuation is no more
special than others once chiral symmetry is explicitly
broken%
\footnote{In studying the singularity of CEP,
{\em approximate} chiral symmetry is at most secondary.
The size of the critical region of CEP and the distance from
TCP are different, interesting issues\cite{HI03}.}.

In this report
we point out that the most important property of CEP
is the divergent fluctuations of the conserved quantities such
as the baryon number and the energy\cite{FO03}.
Since only the hydrodynamic mode contributes to
the susceptibilities of these conserved quantities,
the soft mode associated with CEP should have such a 
property and causes various divergences at CEP through the mixing.
Within the NJL model calculation and the TDGL approach we show that this is indeed the case.

\section{Susceptibility as a spectral sum and its constraints}

Divergence of the susceptibility implies that there appears
at least one soft mode associated with the criticality.
Since the susceptibility $\chi$ is obtained
as a $q$--limit of the
response function $\chi(\omega,\bq)$,
which is analytic in the upper half plane of complex~$\omega$,
one can express the susceptibility as a sum of the
spectral strength over all frequencies:
\begin{eqnarray}
\chi = \chi(0,\bzero^+) 
     = \lim_{\bq \to 0} \chi(0,\bq) 
     = \lim_{\bq \to 0} {{\cal P} \over \pi} \int_{-\infty}^{\infty}
       {d\omega' \over \omega'} \Imag\chi(\omega',\bq) ,
\label{eq:qlimit}
\end{eqnarray}
where a ultra--violet regularization
is understood if necessary.
Spectral enhancement at vanishing $\omega$ is required to get
the critical divergence because the 
spectral function $\rho(\omega,\bq)= 2 {\rm Im}\chi(\omega,\bq)$
itself is usually integrable.

Let us discuss the constraints on the 
spectral functions.
First, in the second--order chiral transition
with the massless quarks,
the sigma mode becomes gapless 
so as to realize the manifest chiral symmetry in the spectrum
together with the pion mode
(see Ref.~\citen{MH03} for another possibility).
In the typical NJL calculation\cite{HK85},
the spectral peak of the sigma meson 
($\bq=0$) grows
sharply and moves to $\omega=0$
as the critical point is approached from the symmetric phase,
which results in the divergent scalar susceptibility.
At CEP with the finite quark mass,
however,
there is no symmetry reason to expect the gapless sigma meson.

The second constraint is a consequence of the conservation of
the baryon number density and the energy density.
The fluctuations of the conserved quantities are 
natural slow modes whose frequencies vanish in 
the long wavelength limit, and constitute a basis of 
hydrodynamics.
For (e.g.,) 
the quark number density $\rho$
there is a current ${\mbf j}$ such that
$\omega \rho - \bq \cdot {\mbf j}=0$,
which implies that $\omega \to 0$ as $\bq \to \bzero$.
A rephrasing of this fact is  that
the spectral function of the conserved density fluctuation
is proportional to the delta function $\omega\delta(\omega)$ in
the $\bq \to 0$ limit%
\cite{HK94,FO03,CMT03}.
We call the spectrum with this property ``hydrodynamic''
in this report.

At CEP the quark number and entropy susceptibilities are
divergent, to which only hydrodynamic spectrum contributes.
This fact directly requires extra softness or critical slowing
of a hydrodynamic mode near CEP to give these divergences.

A remark here may be in order. 
The coupling with the hydrodynamic mode
is the origin of the different
limiting values of the response function at
$(\omega,\bq)=(0,\bzero)$%
\cite{HF03}.
In fact, the $\omega$--limit of $\chi(\omega,\bq)$
\begin{eqnarray}
\chi(0^+,\bzero) &=& 
      \lim_{\omega \to 0} {{\cal P} \over \pi}
       \int_{-\infty}^{\infty}
       {d\omega' \over \omega'-\omega} \Imag\chi(\omega',\bzero)
\label{eq:wlimit}
\end{eqnarray}
does not contain the contribution of the hydrodynamic mode,
which is apparent from a formal relation $\omega \delta(\omega)=0$.

\section{Nambu--Jona-Lasinio model\cite{HF03}}

\begin{figure}[tb]
\centerline{\epsfxsize=0.45\textwidth
\epsffile{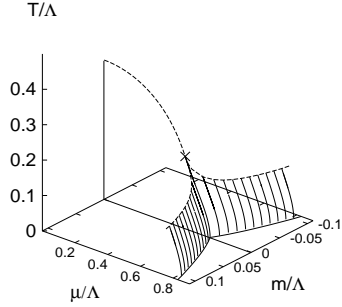}}
\caption{Phase diagram of the NJL model in $T$--$\mu$--$m$ space. 
Three critical (dashed) lines meet at
the tricritical point $\times$.
The phase boundary (hatched) of the first-order transition forms
a wing--like structure.}
\label{fig:phase}
\end{figure}

\subsection{Phase diagram}
We study
the simple Nambu--Jona-Lasinio model\cite{HK94} 
${\cal L}=\bar q (i\!\!\!\not\! \partial - m)q
+g [(\bar q q)^2 +
(\bar q i \gamma_5 \tau^a q)^2] $
defined with the three-momentum cutoff $\Lambda$ and
with the coupling strength $g \Lambda^2=2.5$,
within the mean field approximation. 
The effective potential for the scalar density
$\sigma$ yields
\begin{eqnarray}
\Omega(T,\mu,m;\sigma) &=&
-\nu \int {d^3 k \over (2\pi)^3}
[E_\bk-T \ln (1-n_+)-T \ln (1-n_-)]
+{1\over 4g}\sigma^2,
\label{eq:pressure}
\end{eqnarray}
where $n_\pm = (e^{\beta (E\mp\mu)}+1)^{-1}$, 
$E_\bk=\sqrt{M^2+\bk^2}$, $M=m-\sigma$ and 
$\nu=2 N_f N_c=2 \cdot 2\cdot 3$.

The phase diagram in the $T$--$\mu$--$m$ space is shown
Fig.~\ref{fig:phase}. 
Within the symmetry plane ($m=0$), the chiral broken phase
is separated from the symmetric phase by a critical line,
on which the sigma meson mode is gapless.
At temperatures below the {\it tricritical} point (TCP)%
\footnote{Historically the point $\times$ is called 
tricritical because
three critical lines meet there.}
the boundary becomes the first--order line of
three--phase coexistence\cite{MS03}.
As the explicit breaking term $m\ne 0$ is introduced and increased,
this first--order line is lifted to form a wing--like structure.
The densities of the scalar, baryon number and
entropy are discontinuous across this wing surface in general,
and their susceptibilities contain
 the critical fluctuations at the edge of the wing%
\footnote{
In the vicinity of the critical lines except for TCP, the mean field
approximation is known to break down due to large 
fluctuations. Furthermore, in the chiral broken phase with $m=0$
the pionic fluctuation
is the most important. We neglected these
fluctuations in this work.
}.

\subsection{Spectrum in the scalar channel}

The collective motion of this model
is generated through the (pseudo--) scalar interactions.
The response functions in the scalar and vector channels are easily
evaluated at the RPA level $(a,b=\mu, m)$:
\begin{eqnarray}
\chi_{ab}(iq_4,\bq)
&=&
\Pi_{ab}(iq_4,\bq)+\Pi_{a m}(iq_4,\bq)
{1 \over 1-2g \Pi_{m m}(iq_4,\bq)} 2g \Pi_{m b}(iq_4,\bq),
\label{eq:response}
\end{eqnarray}
where $\Pi_{ab}(iq_4,\bq)$ are the one--loop polarization functions
calculated in the imaginary--time formalism. The real--time
response function is obtained by the usual replacement
$i q_4\to q_0+i\varepsilon$ in the final expression.

\begin{figure}[tb]
\centerline{
\epsfxsize=0.5\textwidth
\epsffile{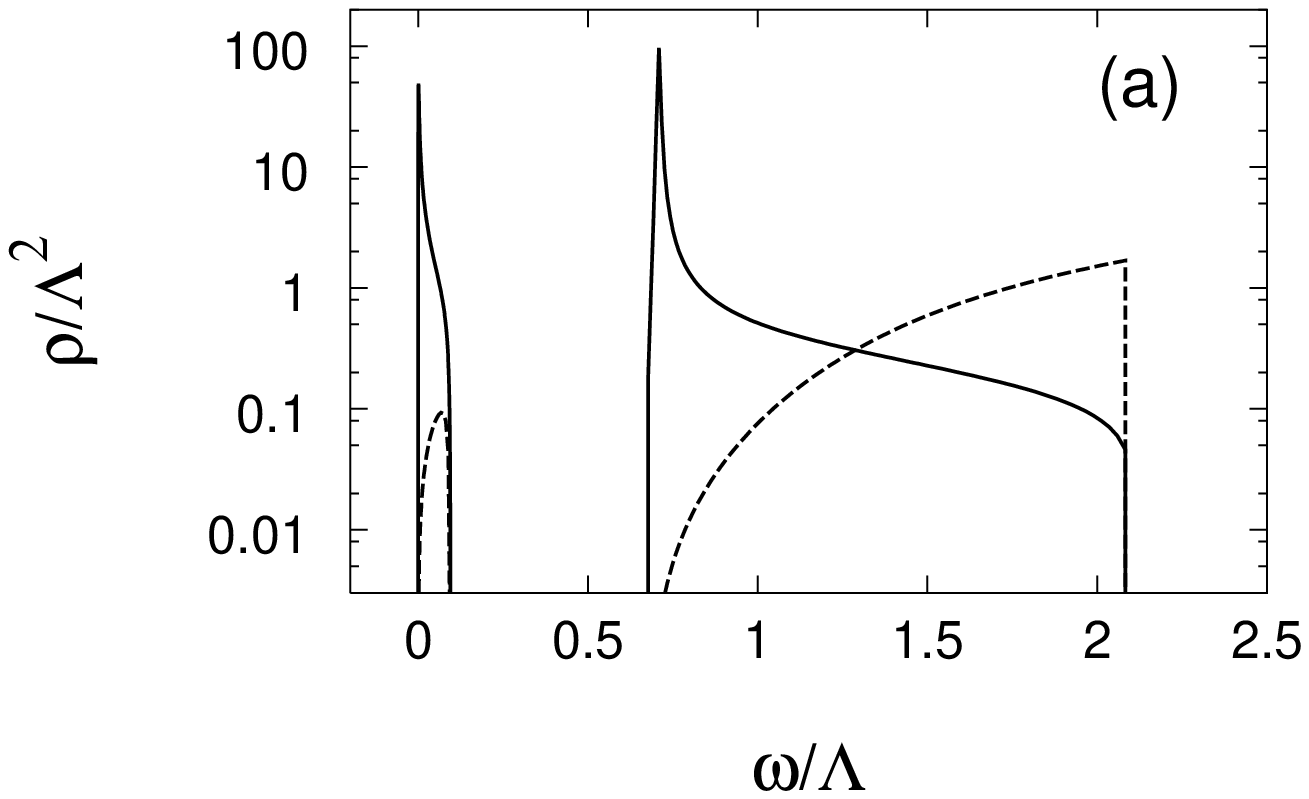}
\hfill
\epsfxsize=0.3\textwidth
\epsffile{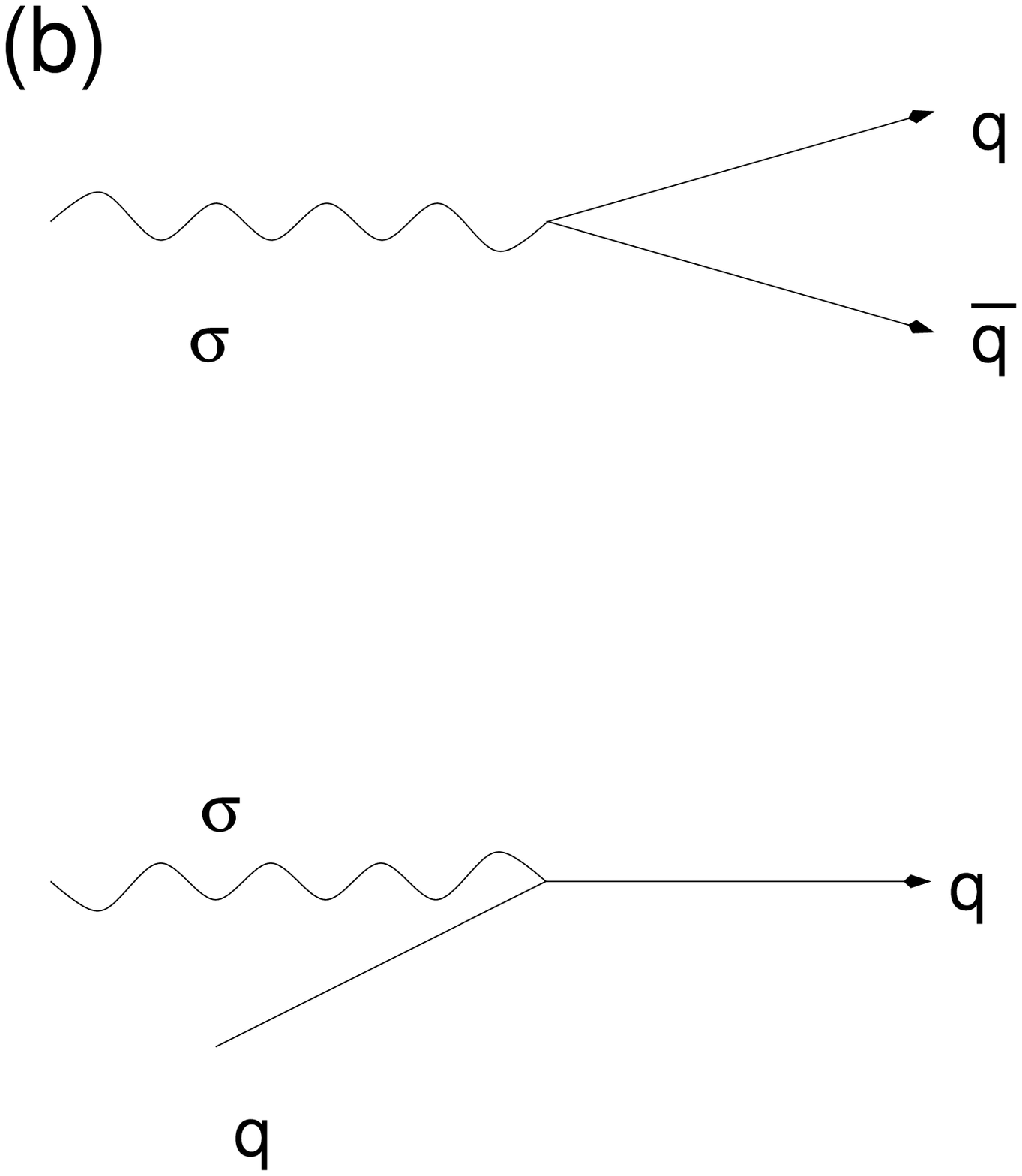}
\hfill}
\caption{(a) Spectral function in the scalar channel (solid) with
$|\bq|/\Lambda =0.1$ at a CEP with $m/\Lambda=0.01$. 
The free gas spectrum (dashed) is also shown for reference.
(b) Typical processes contributing to the spectrum.}
\label{fig:spfunc}
\end{figure}

Let us first consider the scalar channel, in which
collective mode is generated by
the denominator in Eq.~(\ref{eq:response}).
In Fig.~\ref{fig:spfunc} (a) we present the
spectral function
$\rho(\omega,\bq)=2 \Imag \chi_{mm}(\omega,\bq)$ with
$|\bq|/\Lambda=0.1$ at a CEP with $m/\Lambda=0.01$.
We clearly see a peak around $2M$ with $M/\Lambda=0.331 (\propto m^{1/5})$,
corresponding to the sigma meson,
which is interpreted as a collective excitation of the $\bar q q$ pairs
(see Fig.~\ref{fig:spfunc} (b)).
This sigma meson peak cannot give
the critical divergence at CEP because it stays massive there.
The peak is unphysically narrow in our model
because we neglected the $\sigma \leftrightarrow \pi\pi$ coupling,
which if included will broaden the width and decrease the threshold to 
$\sqrt{4m_\pi^2+\bq^2}$ with $m_\pi \sim m^{2/5} \ne 0$. 
The above conclusion will be unaltered in spite of this improvement.

\begin{figure}[tb]
\begin{minipage}{0.48\textwidth}
\centerline{
\epsfxsize=\textwidth
\epsffile{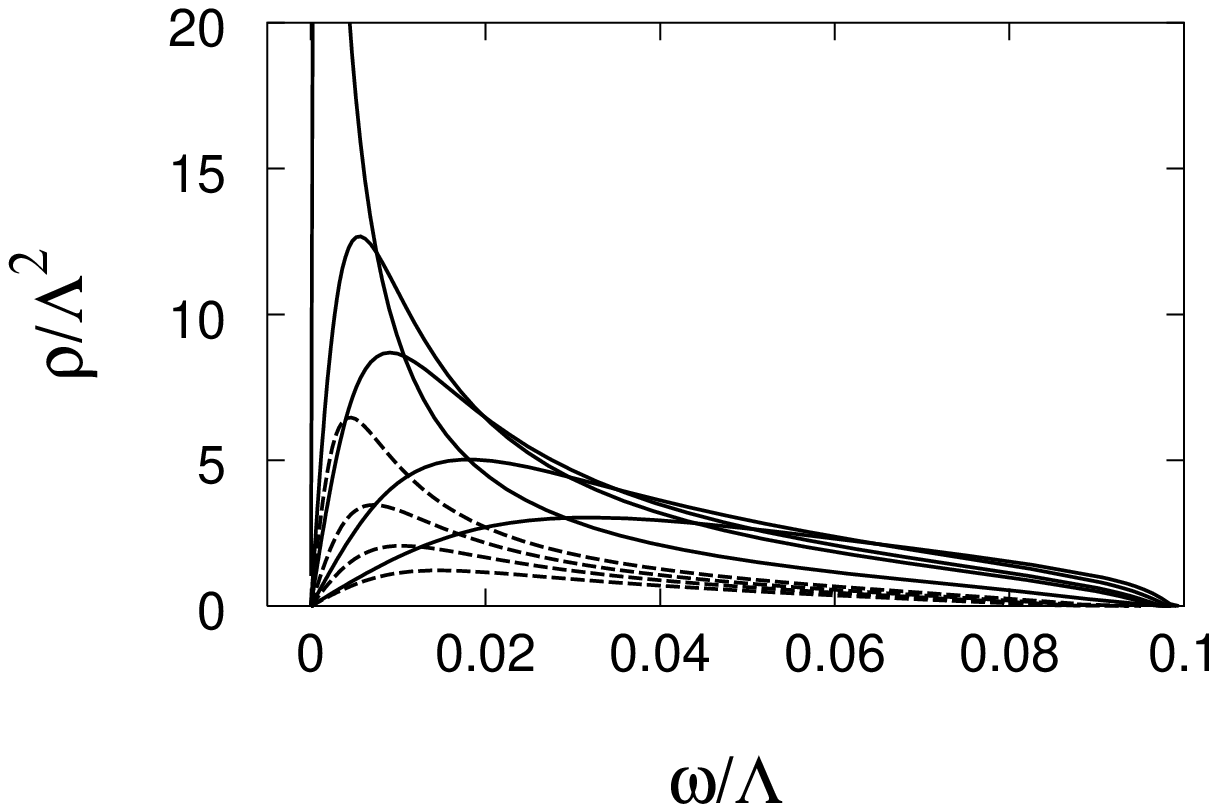}}
\caption{Spectral function in the spacelike momentum region with
$|\bq|/\Lambda=0.1$,
$T=T_c$ and $m/\Lambda=0.01$ for several $\mu$ (see text).}
\label{fig:enhance}
\end{minipage}
\hfill
\begin{minipage}{0.48\textwidth}
\centerline{
\epsfxsize=\textwidth
\epsffile{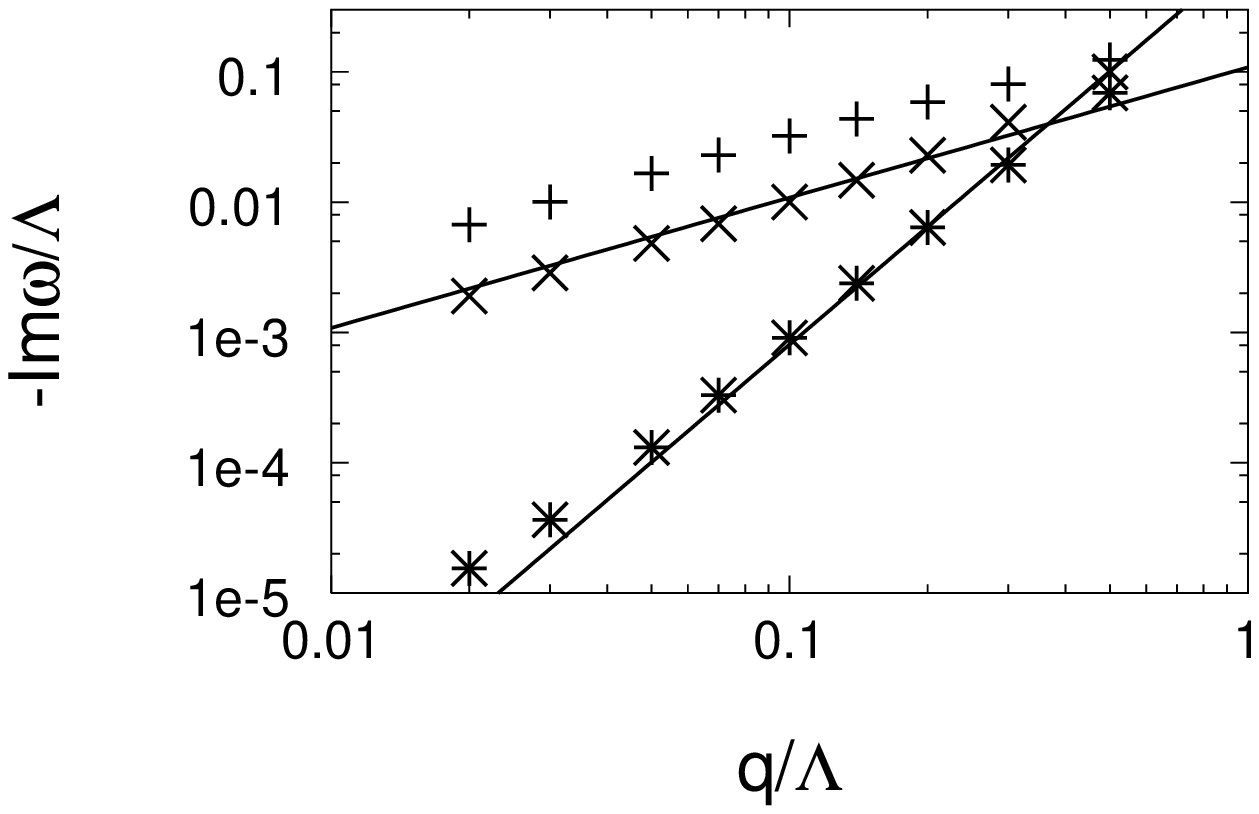}}
\caption{The pole position of $\chi_{mm}(\omega,\bq)$ as a function of
$|\bq|/\Lambda$ for $\mu/\Lambda=$0.56 ($\times$),
$\mu_c/\Lambda$ ($*$), 0.58 ($+$).}
\label{fig:pole}
\end{minipage}
\end{figure}

There is another spectral peak in the spacelike--momentum region
($\omega<|\bq|$), which
is generated by the collective absorption/emission of the scalar fluctuation
by a quark or an anti--quark in medium
(see Fig.~\ref{fig:spfunc} (b)). 
This type of motion is possible only in medium and 
called particle--hole (p--h) excitation, especially at zero temperature.
The excitation energy of this motion is kinematically soft; $\omega \sim
E_{\bk+\bq}-E_\bk\sim  \bv \cdot \bq$.
The peak is strongly  enhanced
as compared with the free gas spectrum shown in the dashed line
in Fig.~\ref{fig:spfunc} (a).
The enhancement of this spectrum is demonstrated for 
$\mu/\Lambda$=0.58, 0.575, 0.572, 0.571, 0.5701=$\mu_c/\Lambda$ (solid lines),
0.5682, 0.5652, 0.56, 0.55 (dashed lines) 
with fixed temperature $T_c/\Lambda=0.19485$
in Fig.~\ref{fig:enhance}.

We found numerically a pole of $\chi_{mm}(\omega,\bq)$
on the negative imaginary axis
of $\omega$, which we expect is associated with this spectral enhancement.
In fact, we confirmed that near CEP 
$\chi_{mm}(\omega,\bq)$ is well described with this single pole
as
\begin{eqnarray}
\chi_{mm}(\omega, \bq) \sim
{1 \over -i {\omega \over \lambda(\bq)}+\chi^{-1}_{mm}(\bq)}
=
{\lambda(\bq) \over -i\omega +\omega_c(\bq)}
\label{eq:pole}
\end{eqnarray}
with $\lambda(\bq) \propto |\bq|$ and 
$\omega_c(\bq)=\lambda(\bq)\chi^{-1}_{mm}(\bq)$.
We note that $\omega_c(\bq) \propto |\bq|$ in the non--critical case
while $\omega_c(\bq) \propto |\bq|^3$ at CEP because
$\chi^{-1}_{mm}(\bq) \propto \bq^2$ there (see Fig.~4).

In the NJL model the singularity 
of the baryon number susceptibility $\chi_{\mu\mu}$
at CEP can be
generated again by the
denominator in (\ref{eq:response}).
It is noteworthy that only the p--h collective mode
can mix with $\chi_{\mu\mu}$ through the coupling because 
$\Pi_{m\mu}(\omega>0,\bzero)=0$.
The importance of the mode with the hydrodynamic character is
consistent with the argument given in \S2.

We conclude that the origin of the divergences
at CEP is the softening of this p--h spectrum in the NJL model. 
We think that 
the essence of this understanding 
still remain valid after the inclusion of the critical fluctuations 
beyond the mean field approximation although these effects
are to be elaborated.
If confinement is treated, the p--h spectrum may come from
the nucleons.

\subsection{Spectral contribution along the critical line\cite{FO03}}

Now that we know the dominance of the p--h mode
in $\chi_{mm}$ at CEP, let us examine its contribution
in the chiral transition with $m=0$ because
the p--h mode always exists in medium.
We define ratio $R$
by
\begin{eqnarray}
R={\chi_{mm}(0,\bzero^+)-\chi_{mm}(0^+,\bzero)\over
  \chi_{mm}(0,\bzero^+) },
\label{eq:ratio}
\end{eqnarray}
which measures the importance of the hydrodynamic spectrum
as explained in \S 2.
This ratio for $\chi_{\mu\mu}$ is always unity
due to the spectral property of the conserved quantity.
In Fig.~5, we present the ratio $R$ of the scalar channel
along the critical line
approached from the broken phase (see also Fig.~1). 
It might be a little surprise that
the p--h mode spectrum gives a finite fraction of the 
critical divergence even in the
second--order chiral transition, and it
finally dominates the spectral sum ($R=1$) at TCP, 
where $\chi_{\mu\mu}$ and $\chi_{TT}$ diverge
in addition to $\chi_{mm}$.
This result seems natural from the argument
given in \S 2.

If the critical line is approached from the symmetric phase, on the
other hand, the p--h collective spectrum does not
contribute to $\chi_{mm}$. Thus
the divergence solely comes from the sigma meson mode ($R=0$)
and $\chi_{\mu\mu}=\Pi_{\mu\mu}(0,\bzero^+)$. 
Actually there is no p--h spectral strength in the scalar channel
at $\bq=0$ in the symmetric phase. 
This is understood as follows:
the absorption amplitude of the collective mode with 
momentum $\bq$ by a quark
$q_L(\bk)$ is proportional to a spinor product
$\bar u_R(\bk+\bq)u_L(\bk)\propto |\bq|$ with a
helicity flip.
(Since a massive quark with a definite chirality
has both helicity components, the helicity flip is unnecessary
in the broken phase.)

\begin{figure}[tb]
\centerline{
\epsfxsize=0.45\textwidth
\epsffile{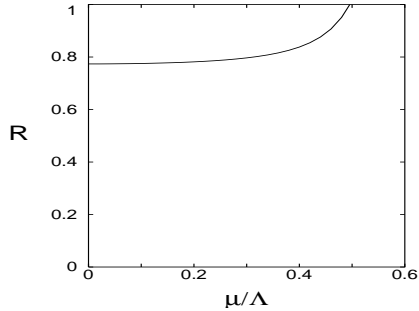}}
\caption{Ratio $R$ along the critical line.}
\end{figure}

\section{Time--dependent Ginzburg--Landau approach\cite{FO03}}

In order to confirm the generality of our result, 
let us repeat our analysis within
the TDGL approach here.
It is essential to introduce another density $\varphi$\cite{HH77}
near TCP which 
is physically a linear combination of the baryon number and the energy
densities, and whose susceptibility diverges at TCP (and at CEP as well).
We expand the GL effective potential  in terms of the ordering densities
$\sigma$ and $\varphi$ around TCP as 
\begin{equation}
\Omega = \int d^3 x \left (
\frac{\kappa}{2}(\nabla \sigma)^2+
a_0 \sigma^2 + b_0 \sigma^4 + c\sigma^6 
+\gamma \sigma^2 \varphi + \frac{1}{2}\varphi^2
-h \sigma -j \varphi  \right ).
\label{eq:GLpot}
\end{equation}
Coupling between $\sigma$ and
$\varphi$ must respect the underlying chiral symmetry.
Eliminating the density $\varphi$ by
$\partial \Omega/\partial \varphi=
\gamma \sigma^2 + \varphi -j=0$,
we recover the usual $\sigma^6$ theory of the free energy 
with $\sigma^2$ and  $\sigma^4$ coefficients replaced by
$a=a_0+\gamma j$ and $b=b_0-\frac{1}{2}\gamma^2$.

\begin{figure}[tb]
\centerline{
\epsfxsize=0.58\textwidth
\epsffile{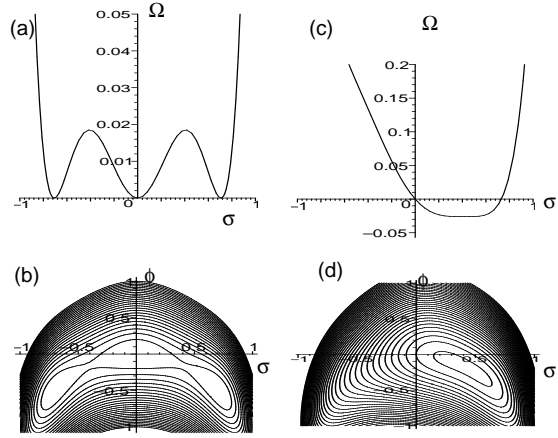}}
\caption{Schematic GL potentials at a three--phase
coexisting point ((a) and (b)) and at a CEP 
((c) and (d)), with single ordering density (upper) and with two ordering
densities (lower).}
\label{fig:GL}
\end{figure}

In the TDGL approach the coefficients are unknown. 
In Fig.~\ref{fig:GL} shown are the typical examples of
the three--phase coexistence ((a), (b)) and the CEP ((c), (d)).
In case of exact chiral symmetry ($h=0$), 
the $\sigma$ direction is obviously
special. 
Although
the GL potential of
single order parameter $\sigma$ becomes flat at CEP,
the actual flat direction is a linear combination 
of $\sigma$ and $\varphi$. 
Both the scalar susceptibility 
$\chi_h=1/(2a+12b \sigma^2 +30 c \sigma^4)$
and the $\varphi$--susceptibility
$\chi_j=1+4 \gamma^2 \sigma^2\chi_h$
diverge there.
We stress here that once $h\ne 0$ one may equally use
an alternative GL potential with single order parameter $\varphi$
defined by eliminating $\sigma$ 
in favor of $\varphi$ in Eq.~(\ref{eq:GLpot})\cite{FO03}.

Differently from the NJL case where the collective
modes are generated by the microscopic interaction and
then coupled with the scalar and vector channels,
the dynamics here in the TDGL approach 
is introduced phenomenologically from
the outset\cite{HH77} via
\begin{eqnarray}
L_\sigma(i\partial_t)\sigma
&=&- \frac{\delta \Omega}{\delta \sigma},\quad
L_\varphi(i\partial_t) \varphi
=- \frac{\delta \Omega}{\delta \varphi}.
\label{eq:TDGLeq}
\end{eqnarray}
One stringent condition on the time dependence is that the density $\varphi$
is a conserved quantity, and possible form of $L_\varphi$ will be
of diffusion $L_\varphi(\omega)=-i \omega / \lambda \bq^2$
or sound--like
$L_\varphi(\omega)=- \omega^2 / \lambda \bq^2$.
On the other hand, it is natural to expect 
$L_\sigma(\omega)=-\omega^2/\Gamma$ (propagation)
or $-i\omega/\Gamma$ (relaxation).
Here we assume coefficients $\lambda$, $\Gamma$ are constant.

By linearizing the equation of motion around the equilibrium densities,
we obtain the condition for the normal mode
\begin{eqnarray}
\det
\left (
\begin{array}{cc}
L_\sigma(i\partial_t)+\Omega_{\sigma\sigma}    &
 \Omega_{\sigma\varphi} \\
\Omega_{\sigma\varphi}     &
L_\varphi(i\partial_t)+\Omega_{\varphi\varphi}
\end{array} \right )
=0,
\label{eq:TDGLmode}
\end{eqnarray}
where $\Omega_{ab}=\delta^2 \Omega/\delta a \delta b|_{\rm eq}$.
Because of  the soft nature of the conserved densities,
the solution for the mixing between ({\em e.g.,}) the propagation and the
diffusion is found in the small--$\bq^2$ limit 
as
\begin{eqnarray}
{-\omega^2 \over \Gamma} &=& -  (\chi_h^{-1}+4 \gamma^2 \sigma^2 ),
\quad
{-i\omega \over \lambda \bq^2}= -  \chi_j^{-1}.
\label{eq:mode}
\end{eqnarray}
The soft mode becomes softer in the mixing among the modes,
and the normal mode with diffusion--like character shows
the critical slowing ($\propto \chi_j^{-1}\to 0$)
while the propagating mode does not at CEP ($\sigma \ne 0$).

The response function is the inverse of Eq.~(\ref{eq:TDGLmode})
with the retarded boundary condition. 
Using the spectral function we can again discuss the
mode contributions to the susceptibilities. 
For the scalar channel, the response function yields
\begin{eqnarray}
\chi_h(\omega, \bq) = 
{L_\varphi(\omega)+\Omega_{\varphi\varphi\bq}
 \over
( L_\sigma(\omega+i\varepsilon)+\Omega_{\sigma\sigma\bq})
(L_\varphi(\omega)+\Omega_{\varphi\varphi\bq})
- \Omega_{\sigma\varphi\bq}^2} .
\end{eqnarray}
The susceptibility 
$\chi_h=\lim_{\bq \to 0}\chi_h(0, \bq)$ is expressed as a spectral sum
of the two modes as
\begin{equation}
\chi_h =\chi_h  
\left (
{\chi_h^{-1} \over \chi_h^{-1}+4 \gamma^2 \sigma^2}
+ 
{4 \gamma^2 \sigma^2 \over \chi_h^{-1}+4 \gamma^2 \sigma^2}
\right ),
\end{equation}
where the first term in the bracket is from the propagating mode
and the second from the diffusion--like mode in Eq.~(\ref{eq:mode}).
From this simple expression we find that
in the chirally symmetric phase ($\sigma=0$) there is no contribution
from the diffusion mode $R=0$. In the second order chiral transition
approached from the broken phase, a finite fraction of the
divergence comes from the diffusion--like mode ($0<R<1$), and
$R=1$ at TCP. In the criticality of CEP the diffusion--like
mode saturates the divergence ($R=1$)
because $\chi_h^{-1}=0$ and $\sigma\ne 0$.
Incidentally,
in the $\omega$--limit the diffusion--like contribution drops out
due to its hydrodynamic nature. These discussions are completely in
parallel to the NJL case, and consistent with the argument in
\S 2.

\section*{Acknowledgments}
The authors are grateful to the organizers
for the opportunity of the presentation and 
the stimulating discussions at the workshop. They also thank K.~Ohnishi
for the discussions in the early stage of this work.
This work is supported in part by  the Grants-in-Aid for Scientific Research 
of Monbu-kagaku-sho (No.\ 13440067).

\end{document}